\documentclass[11pt,letter]{article}

\usepackage{amsmath, amsthm, amssymb}
\usepackage{amssymb,epsfig}

\newcommand{\bx}{\mathbf{x}}

\makeatletter

\gdef\@punct{.\ \ }  
\def\@sect#1#2#3#4#5#6[#7]#8{%
  \ifnum #2>\c@secnumdepth
     \def\@svsec{}
  \else
     \refstepcounter{#1}\edef\@svsec{%
     \ifnum #2>0{{\csname the#1\endcsname}}.\fi%
    \hskip .5em}
  \fi
  \@tempskipa #5\relax
  \ifdim \@tempskipa>\z@
     \begingroup #6\relax
       \@hangfrom{\hskip #3\relax\@svsec}{\interlinepenalty \@M #8\par}
     \endgroup
     \csname #1mark\endcsname{#7}
     \addcontentsline{toc}{#1}{\ifnum #2>\c@secnumdepth\else
          \protect\numberline{\csname the#1\endcsname}\fi#7}
  \else
     \def\@svsechd{#6\hskip #3\@svsec #8\@punct\csname
#1mark\endcsname{#7}
     \addcontentsline{toc}{#1}{\ifnum #2>\c@secnumdepth \else
          \protect\numberline{\csname the#1\endcsname}\fi#7}}
  \fi
  \@xsect{#5}}

\def\@ssect#1#2#3#4#5{\@tempskipa #3\relax
  \ifdim \@tempskipa>\z@
     \begingroup #4\@hangfrom{\hskip #1}{\interlinepenalty \@M
#5\par}\endgroup
  \else \def\@svsechd{#4\hskip #1\relax #5\@punct}\fi
  \@xsect{#3}}

\newcommand{\imp}{\mathop{\mathrm{imp}}\nolimits}
\newcommand{\row}{\mathop{\mathrm{row}}\nolimits}
\newcommand{\clique}{\mathop{\mathrm{clique}}\nolimits}
\newcommand{\degree}{\mathop{\mathrm{degree}}\nolimits}
\newcommand{\mixed}{\mathop{\mathrm{mixed}}\nolimits}
\def\qed{\hskip 3pt \hbox{\vrule width4pt depth2pt height6pt}}

\newtheorem{Lemma}{Lemma}

\newtheorem{Theorem}[Lemma]{Theorem}
\newtheorem{Proposition}[Lemma]{Proposition}
\newtheorem{Corollary}[Lemma]{Corollary}
\newtheorem{Definition}[Lemma]{Definition}

\begin{document}
\title{Performance of distributed mechanisms for flow admission in wireless adhoc networks}

\author{Ashwin~Ganesan%
  \thanks{This work was carried out while the author was with the ECE Department, University of Wisconsin at Madison, WI 53706,
USA. This work was presented in part at the International Conference on Networks and Communications (GraphHoc and NetCoM), Chennai, India, December 2009 \cite{Ganesan:2009}, and at the International Conference on Recent Trends in Graph Theory and Combinatorics (an ICM Satellite Conference), Cochin, India, August 2010 \cite{Ganesan:ICRTGC}.  This paper is a revised version of the Technical Report \cite{Ganesan:2008}. The author is with the Department of Mathematics, Amrita School of Engineering, Amrita University, Coimbatore, Tamil Nadu, India. Correspondence address: \texttt{ashwin.ganesan@gmail.com}.
}
}

\date{}

\maketitle

\vspace{-.5cm}

\begin{abstract}
\noindent  Given a wireless network where some pairs of communication links interfere with each other, we study sufficient conditions for determining whether a given set of minimum bandwidth quality-of-service (QoS) requirements can be satisfied.  We are especially interested in algorithms which have low communication overhead and low processing complexity.  The interference in the network is modeled using a conflict graph whose vertices correspond to the communication links in the network.  Two links are adjacent in this graph if and only if they interfere with each other due to being in the same vicinity and hence cannot be simultaneously active.  The problem of scheduling the transmission of the various links is then essentially a fractional, weighted vertex coloring problem, for which upper bounds on the fractional chromatic number are sought using only localized information.  We recall some distributed algorithms for this problem, and then assess their worst-case performance.  Our results on this fundamental problem imply that for some well known classes of networks and interference models, the performance of these distributed algorithms is within a bounded factor away from that of an optimal, centralized algorithm.  The performance bounds are simple expressions in terms of graph invariants.  It is seen that the induced star number of a network plays an important role in the design and performance of such networks.

\end{abstract}




\section{Introduction}

In recent years there has been an increasing interest in using data networks to support a wide variety of applications, each requiring a different Quality-of-Service (QoS).  For example, real-time applications such as voice, video and industrial control are time-sensitive and require that the delay be small, while for other data applications the sender may require that a constant, minimum bit-rate service be provided.  In the simplest and lowest level of service, such as the one provided in the Internet Protocol service model, the network makes a best-effort to deliver data from the source to destination, but it makes no guarantees of any kind, so it is possible that packets can get dropped, delayed, or delivered out of order. However, this basic level of service is insufficient for many data applications such as video-conferencing that also have a minimum bandwidth requirement.  We consider in this work applications requiring a minimum bandwidth quality-of-service.

\bigskip \noindent Consider a wireless communication network where nodes (which represent wireless devices such as laptops, phones, routers, sensors, etc) wish to communicate with each other using a shared wireless medium.  Any given pair of nodes may make a request for a dedicated point-to-point link between them that supports their required bit-rate quality-of-service.  The objective of the \emph{admission control} mechanism is to decide whether the desired service can be provided, given the available resources, without disrupting the service guaranteed to previously admitted requests.  This mechanism needs to take into account the fact that nodes in the same vicinity contend for the shared wireless medium and hence can cause interference effects if simultaneously active. Also, for reasons such as low communication overhead and scalability, it is desired that this decision be made in some decentralized fashion.  Thus, each node may have access only to information pertaining to its local neighborhood and not about the entire communication network.  These two requirements - that the decision take into account \emph{interference} due to neighboring nodes and that it be made in a \emph{decentralized} or distributed manner - are crucial aspects of the particular problem we study.  If a decision to admit a request is made, the \emph{scheduling} problem is to schedule the transmissions of the various nodes so as to provide the service level that was guaranteed.  The focus of most of our work here is on the admission control problem and not the scheduling problem.  For an introduction to the flow admission control problem, see \cite{Bertsekas:Gallager:1992}.

\bigskip \noindent More formally, the wireless network model and desired QoS are specified as follows.
Let V be a set of nodes and $L \subseteq V \times V$ be a set of
communication links.   Each link $(i,j) \in L$ makes a demand to
transmit information from $i$ to $j$ at a rate of $f(u,v)$ b/s.  The
total bandwidth of the shared wireless medium available for the communication network $G=(V,L)$ is $C$~b/s. The main problem studied here is to determine whether a set of demands $(f(\ell):\ell \in
L)$ can be satisfied.  Of course, if all the links can be simultaneously
active, the set of demands $(f(\ell): \ell \in L)$ can be satisfied as long
as each individual demand is at most $C$.  However, due to
interference effects nodes in the same vicinity contend for the shared
wireless medium and hence cannot be active at the same time.
For example, in IEEE 802.11 MAC
protocol-based networks, any nodes adjacent to node $i$ or to node
$j$ are required to be idle while the communication $(i,j)$ takes
place. (The reason for this constraint is that when the
communication from $i$ to $j$ takes place, $j$ also sends
acknowledgement of the receipt of information back to $i$, so nodes
adjacent to $j$ cannot receive information from other nodes during
this time due to interference.)
We will use a more general interference model which includes
this special case and which is still tractable enough for further study.

\bigskip \noindent The interference in the network is modeled using a \emph{conflict graph}.  Given a network graph
$G = (V,L)$, define its conflict graph to be $G_C = (L,L')$, where
two links $\ell_1$ and $\ell_2$ are adjacent in the conflict graph if and
only if they interfere with each other due to being in the same vicinity and hence cannot be simultaneously active.  This interference
model has been studied recently by a number of authors; for example,
see Jain et al \cite{Jain:Padhye:etal:03}, Hamdaoui and Ramanathan
\cite{Hamdaoui:Ramanathan:05}, and  Gupta, Musacchio and Walrand
\cite{Gupta:Musacchio:Walrand:07}.  A special case of this model,
where two links are considered to be interfering if and only if they
are incident (in the network graph) to a common node, has been studied earlier by Hajek
\cite{Hajek:1984}, Hajek and Sasaki \cite{Hajek:Sasaki:1988}, and
Kodialam and Nadagopal \cite{Kodialam:Nandagopal:2005}.   While we
will study the more general model, it still does not incorporate
parameters such as the actual signal-to-interference-and-noise ratio
at the receiver node in its decision making process.  In our work
below we will also revisit the special case just mentioned
and obtain some
new results in the distributed setting.

\bigskip \noindent The notion of a conflict or interference graph was introduced earlier, outside the context of wireless networks, by Chaitin \cite{Chaitin:1982} for solving the register allocation problem.  A register is a high-speed memory location on a computer's CPU and the register set is usually of a very limited size and much faster compared to ordinary memory.  During the register allocation phase of compiling, a decision needs to be made as to which values to keep in registers and which to keep in memory at each point in the object code.  One approach is for the compiler to construct a conflict graph, where the vertices of this graph correspond to values, and there is an edge between two vertices if and only if the two values are simultaneously live at some point in the code.  A valid coloring of a graph is an assignment of one color to each vertex so that adjacent nodes are assigned different colors.  If it is not possible to color the conflict graph using $r$ colors, where $r$ is the number of registers available for use, then one of the nodes (variables) is moved to memory and deleted from the conflict graph, and a valid coloring is again attempted.  Many commercial and research compilers use a coloring approach on the conflict graph during the register allocation phase.

\bigskip \noindent In the context of wireless networks, the notion of coloring the conflict graph has been used to study the spectrum allocation problem in cellular communications \cite{Hale:1980}.  In this setting, the vertices of the conflict graph correspond to base stations.  Two vertices are adjacent iff the corresponding base stations are close enough to interfere while using the same frequency.  Each base station represents its area of service (such as a hexagonal region in the plane) and makes a demand for a certain number of frequencies which is proportional to the traffic demand for that region.  The frequency assignment problem is to assign a set of  frequencies (colors) to each base station, so that the demand for each base station is met, so that interfering base stations are assigned nonoverlapping sets, and so that the minimum number of frequencies is used.  This problem is essentially equivalent to a problem sometimes referred to as the weighted vertex coloring problem.

\bigskip \noindent In the admission control problem studied in our work, the vertices of the conflict graph represent the links in the communication network.  The quality-of-service metric is specified in terms of the bandwidth desired by each link.  This gives rise to one demand value for each vertex of the conflict graph, and this value could possibly be non-integral.  The admission control problem is then to determine whether these demands can be satisfied using a specified amount of resource (total available bandwidth).   This problem is different from the classical weighted vertex coloring problem in some ways.  First, fractional solutions to the coloring problem are also admissible, as indicated by the linear programming formulation given below. Second, our emphasis is on decisions that can be made in a decentralized manner, i.e. using only localized information.  Finally, the conflict graph sometimes has additional structure derived from the structure of the links in the network or the interference model.   The admission control problem studied here is essentially that of obtaining, using only localized information, an upper bound on the amount of resource required to satisfy a demand pattern. The scheduling problem, which we do not study here, concerns how these resources are actually allocated or managed.

\bigskip \noindent This paper is organized as follows.  The admission control problem studied here is formulated precisely in Section~\ref{sec:model:problem:formulation}.  Decentralized solutions to this problem are then studied:  Section~\ref{sec:row:constraints} is on the row constraints, Section~\ref{sec:degree:mixed} is on the degree and mixed conditions, and Section~\ref{sec:clique:constraints} is on the clique constraints.  These results provide sufficient conditions and distributed algorithms for admission control.  The main new results here concern the worst-case performance of these algorithms.   Finally, in Section~\ref{section:examples} the results obtained thus far are applied to some specific examples such as unit disk networks and networks with primary interference constraints.  In order to keep this paper as self-contained and accessible as possible, we recall along the way some of the known results and their proofs from the literature.

\subsection{Model and problem formulation}
\label{sec:model:problem:formulation}
We first state the flow control problem
formally. Then, in order to avoid repeating trivialities throughout the
paper, we will present an equivalent reformulation of the problem that
ignores many of the
constants and variables and involves just the essential details. We
will work only with this reformulation in the rest of this paper.

\bigskip \noindent Let $G=(V,L)$ be a network graph, where $L \subseteq V \times V$.
Each link $\ell \in L$ has a maximum transmission capacity of $C_\ell$
b/s, and there is a demand to use that link at some rate $f(\ell)$ b/s.  The
total available bandwidth of the shared wireless medium is $C$ b/s.
The conflict graph $G_C=(L,L')$ specifies which pairs of links
interfere with each other: two links are adjacent in the conflict graph iff they interfere with each other when they are simultaneously active.  The main problem we study is to determine, using only localized information, whether the set of demands $(f(\ell): \ell \in L)$
can be satisfied.  More precisely, an independent set of a graph
$G_C=(L,L')$ is a subset $I \subseteq L$ of elements that are pairwise nonadjacent.  If the set of links that are simultaneously active is an independent set, then these links cause no interference with each other and can have (a part of) their demands satisfied during the same time slot.
 Let  $\mathcal{I}(G_C)$ denote the set
of all independent sets of $G_C$. [Note that this set can grow exponentially with
the size of the graph. For example, a graph on $n$ vertices consisting of $n/3$ disjoint triangles has $3^{n/3} \approx 1.4^n$ maximal independent sets.]
 A \emph{schedule} is a map $t:\mathcal{I}(G_c) \rightarrow
 \mathbb{R}_{\ge 0}$.  The schedule assigns to each independent set
 $I_j$ a time duration $t_j = t(I_j)$, which specifies the fraction of
 time that  the links in $I_j$ are active.  A schedule $t$ is said to satisfy a set
 of demands $(f(\ell): \ell \in L)$ if, for each $\ell \in L$, $\sum_{I_j:\ell \in
   I_j} t_j C_l \ge f(\ell)$, and if the duration of the schedule
 $\sum_{I_j} t_j$ is at most 1.  A schedule is said to be optimal if
 it satisfies the demand of all the links and has minimum duration.

\bigskip \noindent A schedule $t$, as defined above, can be implemented as follows.
We construct a periodic schedule with period $T$ seconds, say, where
we can choose $T$ arbitrarily. Then divide the time axis into frames of
duration $T$, and further subdivide each frame into subintervals of
length $t_jT$.  The schedule is implemented by letting the set of
links $I_j$  be active during the subinterval of duration $t_jT$.
During this subinterval each link $\ell
\in I_j$ transmits at its maximum transmission rate $C_\ell$. It is
 seen that if the schedule $t$ satisfies the demand vector, then so does this
implementation.   We shall assume throughout that the entries of vectors $(C_\ell)$ and $(f(\ell))$ are all rational
numbers.  Since the linear program defining the duration of an optimal schedule has a rational optimal solution, it can be assumed that the schedule $t$ is a map into the rationals. Let $K$ be the least common multiple of
the denominators of $t_j$.  Then,  divide the time axis into frames of
duration $T$, and further subdivide each frame into $K$ subintervals of
equal length.  Suppose $M_j=K t_j$; then in each frame, we let the
links $I_j$ be active for $M_j$ time slots; we let each
link $\ell \in I_j$ transmit at its maximum transmission rate $C_\ell$
during each time slot when it is active.   Such an implementation meets the demands satisfied by the schedule $t$. 

\bigskip \noindent \textbf{Reformulation.} Suppose we are given a network graph
$G=(V,L)$ and a conflict graph $G_C=(L,L')$ that specifies which pairs of links interfere
with each other. Let $\tau(\ell)$ denote the
amount of time when link $\ell$ demands to be
active. A link demand vector $(\tau(\ell): \ell \in L)$
 is said to be feasible within
 time duration $[0,T]$ if
 there exists a schedule of duration at most $T$ that satisfies the
 demands.  We will often assume, for simplicity of exposition, that
 $T=1$.
Note that a schedule is a map  $t:\mathcal{I}(G_c) \rightarrow
 \mathbb{R}_{\ge 0}$ that assigns to each independent set $I_j$ of the
 conflict graph a time
 duration $t(I_j)$. A link $\ell$ is then active for total duration
 $\sum_{j:\ell \in I_j} t(I_j)$.
The admission control problem is to determine whether there exists a schedule of
duration at most $T$ that satisfies the link demand vector $\tau$.  The
scheduling problem is to realize such a schedule.  We are interested in solutions that can be implemented using only localized information and with low processing cost. For simplicity of exposition, we shall assume that $G_C$ is connected; if this is not the case we can work with each connected component separately and the results here still apply.

\bigskip \noindent In a communication network with interference constraints, the problem of determining the minimum duration of a schedule is essentially that of computing the fractional chromatic number of a weighted graph.  This problem, which corresponds to the problem solved by an optimal, centralized algorithm, is NP-hard in the general case \cite{Grotschel:Lovasz:Schrijver:1981}.  The  admission control mechanisms studied here provide upper bounds on the fractional chromatic number; furthermore, these upper bounds have the advantage that they can be computed efficiently and implemented in distributed systems.

\bigskip \noindent \textbf{Notation.}
It will be convenient to use the following notation \cite{Bollobas:1998}. Let $G=(V,E)$ be a simple, undirected graph. For $v \in V$,
$\Gamma(v)$ denotes the neighbors of $v$. $\alpha(G)$ denotes the
maximum number of vertices of $G$ that are pairwise nonadjacent. For
$V' \subseteq V$, $G[V']$ denotes the induced subgraph whose
vertex-set is $V'$ and whose edge-set is those edges of $G$ that
have both endpoints in $V'$. For any $\tau: V \rightarrow \mathbb{R}$
and any $W \subseteq V$, define $\tau(W):= \sum_{v  \in W}
\tau(v)$.

\subsection{Prior work and our contributions}

The take off point for our work is the prior work of \cite{Gupta:Musacchio:Walrand:07} and  \cite{Hamdaoui:Ramanathan:05}.  Their work proves that certain distributed algorithms provide sufficient conditions for admission control.  They call these conditions the row constraints \cite{Gupta:Musacchio:Walrand:07} (or rate condition \cite{Hamdaoui:Ramanathan:05}), the degree condition \cite{Hamdaoui:Ramanathan:05}, mixed condition \cite{Hamdaoui:Ramanathan:05}, and scaled clique constraints \cite{Gupta:Musacchio:Walrand:07}.  Furthermore, for unit disk networks, the scaled clique constraints are shown to be a factor of 2.1 away from optimal \cite{Gupta:Musacchio:Walrand:07}.

\bigskip \noindent The main results of this paper are along the following lines: the exact worst-case performance of these distributed admission control mechanisms is characterized and it is thereby shown that these mechanisms can be arbitrarily far away from optimal; we then show that for some well known classes of networks and interference models, these distributed algorithms are actually within a bounded factor away from optimal.  The classes of networks and interference models we study include unit disk networks and networks with primary interference constraints.

\bigskip \noindent More specifically, we introduce the notion of the induced star number of a graph and show that it determines the exact worse-case performance of the row constraints.  This implies, for example, that for unit disk networks there is a simple and efficient distributed admission control \emph{and} distributed scheduling mechanism which is at most a factor of 5 away from optimal, and for networks with primary interference constraints this mechanism is at most a factor of 2 away from optimal.  The performance of two other sufficient conditions - namely the degree condition and mixed condition - is also studied.   Finally, the results obtained thus far are applied to some specific classes of networks and interference models.  For example, it is seen that  the scaled clique constraints are at most a factor of 1.25 away from optimal for networks with primary interference constraints.  It is also seen that the different sufficient conditions studied here are in general  incomparable (each condition is neither stronger nor weaker than the others).

\bigskip \noindent Our results quantify the performance of some practically-efficient distributed mechanisms used for flow admission control.  The induced star number of a network is seen to be a crucial parameter that effects the performance of such distributed systems, and hence needs to be as close to unity as possible when designing such networks.  It is also seen that the row constraints can consume up to twice the amount of resources compared to the mixed condition for the same set of demands.  Quantitative performance results of this nature assessing the performance of the various distributed algorithms are also obtained.

\section{Row constraints}
\label{sec:row:constraints}
We now present a sufficient condition for flow admission control that can be implemented in a distributed manner. Given a conflict graph $G_C=(L,L')$, link
demand vector $(\tau(\ell):\ell \in L)$ and $T$, a sufficient condition for feasibility is given by the following result  (see \cite[Thm.~1]{Hamdaoui:Ramanathan:05}, \cite[Thm.~1]{Gupta:Musacchio:Walrand:07}):

\begin{Proposition}
\label{prop:row:constraints:sufficient}
If $\tau(\ell) + \tau(\Gamma(\ell)) \le T$ for each $\ell \in L$, then the
demand vector $(\tau(\ell): \ell \in L)$ is feasible within duration $T$ .

\end{Proposition}

\noindent \emph{Proof}: Pick any ordering of the links
$\ell_1,\ell_2,\ldots,\ell_m$. Assign link $\ell_1$ the time interval $J_1 =
[0,\tau(\ell_1)]$, and initialize $J_j=\phi,~j=2,\ldots,m$.  For each
$i=1,\ldots,k$, assume $\ell_i$ has already been assigned time interval
$J_i \subseteq [0,T]$.  Here, $J_i$ need not be one continuous time
interval; it can be a union of disjoint intervals, but its overall
length is $\tau(\ell_i)$.  Since $\tau(\ell_{k+1}) + \tau(\Gamma(\ell_{k+1}))
\le T$, it follows that  $\tau(\ell_{k+1}) + \tau(\Gamma(\ell_{k+1}) \cap
\{\ell_1,\ldots,\ell_k\}) \le T$.  Hence, it is possible to assign to $\ell_{k+1}$ some subset $J_{k+1}
\subseteq [0,T]$  which is nonoverlapping with the
intervals already assigned to the neighbors of $\ell_{k+1}$ and which has duration
$\tau(\ell_{k+1})$. We can repeat this procedure for the remaining links $\ell_{k+2},\ldots,\ell_m$, in turn. \hfill\qed

\bigskip \noindent These constraints are called the row constraints in
\cite{Gupta:Musacchio:Walrand:07} and the rate condition in
\cite{Hamdaoui:Ramanathan:05}.  Let $A=[a_{ij}]$ be the 0-1 valued
$n \times n$ adjacency matrix of $G_C$ where $a_{ij} = 1$ iff $i=j$
or $\ell_i$ and $\ell_j$ are interfering.  Let $\mathbf{1}$ denote the vector whose every entry is 1.  Then the sufficient condition
above is equivalent to the condition $A \tau \le  T\mathbf{1}$ on the
rows of $A$, hence the name \emph{row constraints}.  Since these constraints depend
on the flow rate (the demand value) of the neighboring links, it is
called the rate condition in \cite{Hamdaoui:Ramanathan:05}, where
the authors also study a degree condition which depends only on the
number of neighboring links rather than on their actual demand values.  We chose to use the phrase row constraints (rather than the rate condition) because all these results also apply in a general context  where the variables $\tau(l)$  need not refer to only rate values,  or even just time durations, but to any demand for resources by competing entities, where the competition is modeled by a conflict graph.

\bigskip \noindent The proof of Proposition~\ref{prop:row:constraints:sufficient} also gives
a very efficient algorithm for checking feasibility. It provides
both a distributed admission control mechanism as well as a distributed
scheduling algorithm: when a link $\ell_i$ that is currently inactive
makes a demand to be active for duration $\tau(\ell_i)$, the admission
control mechanism can be implemented efficiently by just checking
the condition above for $\ell_i$ and its neighbors.  The
information required by a link to check this condition is just its
demand and the demand of its neighbors.  Furthermore, the distributed scheduling algorithm that meets the demand for link $\ell_i$  needs to know only the time intervals already
assigned to the neighbors of $\ell_i$ in order to determine the time
interval for $\ell_i$.

\bigskip \noindent It is known (cf.
\cite[Thm.~5]{Hamdaoui:Ramanathan:05}) that this sufficient condition is
also necessary if the conflict graph is the disjoint union of
complete graphs. Indeed, if a given set of links
are pairwise interfering, then the duration of an optimal schedule
for those links is equal to the sum of their individual demands.
It turns out that the converse is also true: the sufficient
condition above becomes necessary only if the conflict graph is the
disjoint union of complete graphs.  To prove this, suppose that the
conflict graph is connected but not complete.  Then there exist
three links $\ell_1,\ell_2$ and $\ell_3$ such that $\ell_1$ and $\ell_2$ interfere,
$\ell_2$ and $\ell_3$ interfere, but $\ell_1$ and $\ell_3$ do not interfere (otherwise, the relation of two links being interfering would be an equivalence relation and the conflict graph would be complete).  The link demand vector $(1 - \varepsilon, \varepsilon,
1-\varepsilon,0,\ldots,0)$ is feasible within duration [0,1] but
does not satisfy the row constraint for $\ell_2$.  This proves the converse. We shall show that this observation -
that the row constraints above are also a necessary condition if and only if the
conflict graph is the disjoint union of complete graphs - can be deduced from a more general result  (cf.  Theorem~\ref{theorem:star:number:row:factor}).

\subsection{Row constraint polytope and induced star number}

In order to quantify how far the row constraints  are from optimal, it
will be useful to introduce its associated polytope.  Given a
conflict graph, let $P_I$ denote its independent set polytope. This
polytope is defined as the convex hull of the characteristic
vectors of the independent sets of the graph.  Note that $P_I$ is exactly equal
to the set of all link demand vectors which are feasible within one
unit of time.  For the given conflict graph, let $P_{\row}$ denote the set of all link demand
vectors that satisfy the row constraints for $T=1$; that is,
$$P_{\row} := \{ \tau \ge 0: \tau(\ell_i) + \tau(\Gamma(\ell_i)) \le 1 ~\forall i\}.$$
Since the row constraints are sufficient, $P_{\row} \subseteq P_I$.
Also, note that $\beta P_{\row} = \{ \tau:
\tau(\ell_i)+\tau(\Gamma(\ell_i)) \le \beta\}$.  Define the scaling
factor $$\beta_{\row} := \inf\{\beta \ge 1: P_I
\subseteq \beta_{\row} P_{\row}  \}.
$$
Equivalently, $$\beta_{\row} = \sup_{\tau \in P_I} \max_i \{
\tau(\ell_i) + \tau(\Gamma(\ell_i))\}.
$$
So $P_{\row} \subseteq P_I \subseteq \beta_{\row}
P_{\row}$, and $\beta_{\row}$ is the smallest scaling factor which converts the sufficient condition into a necessary one.

\bigskip \noindent It has been pointed out (cf. \cite{Gupta:Musacchio:Walrand:07}) that the row constraints can be
arbitrarily far away from optimal. For example, suppose the network
consists of links $\ell_1,\ldots,\ell_{d+1}$, where $\ell_1$ interferes with
each of $\ell_2,\ldots,\ell_{d+1}$ and there is no interference between the remaining links.
Then the conflict graph is a star graph. The link demand vector
$(\varepsilon,1-\varepsilon,\ldots,1-\varepsilon)$ is feasible
within one unit of time, but the row constraint for $\ell_1$ has value
$\tau(\ell_1)+\tau(\Gamma(\ell_1)) = \varepsilon + (1-\varepsilon)d$ which can be
made arbitrarily close to $d$ as $\varepsilon$ approaches 0.  This shows that $\beta_{\row}
\ge d$.  We prove next that the opposite inequality also holds, i.e. the row constraints can be a factor $s$
away from the optimal schedule time for some demand vector
only if the conflict graph contains a star graph on $\lceil s+1 \rceil$
vertices as an induced subgraph.

\begin{Definition}  The \emph{induced star number} of a graph $H$ is
defined by
$$ \sigma(H) := \max_{v \in V(H)} \alpha(H[\Gamma(v)]).\quad\quad\qed$$
\end{Definition}

\bigskip \noindent  Hence, the induced star number of a graph is the number of leaf
vertices in the maximum sized star of the graph.  Note that $\sigma(H)$ equals 0 or 1 iff $H$ is the disjoint union of complete graphs. The induced star number of a graph determines
exactly how close the row constraints are to optimal in the worst case:
\begin{Theorem}
\label{theorem:star:number:row:factor}
 Let $G_C$ be a conflict graph.
The exact worst-case performance of the row constraints is given by $\beta_{\row} = \sigma(G_C)$.
\end{Theorem}

\noindent \emph{Proof:}  It has already been pointed out above
that $\beta_{\row} \ge \sigma(G_C)$.  To prove the opposite
inequality, suppose $\tau \in P_I$ and let $\ell_i$ be any link with
demand $\tau(\ell_i) = \delta \in (0,1)$. It suffices to show that
$\tau(\ell_i) + \tau(\Gamma(\ell_i)) \le \sigma(G_C)$.  Since links in
$\Gamma(\ell_i)$ must be scheduled at time intervals disjoint from
those of $\ell_i$ and since $\tau \in P_I$, there exists a schedule
satisfying the demands of just the links $\Gamma(\ell_i)$ which has
duration at most $1-\delta$. The maximum number of pairwise
non-interfering links in $\Gamma(\ell_i)$ is $\sigma(G_C)$, and so at
most $\sigma(G_C)$ links can be simultaneously active during any time
interval of this period of duration $1-\delta$. Hence,
$\tau(\Gamma(\ell_i)) \le \sigma(G_C)(1-\delta)$.  So $\tau(\ell_i) +
\tau(\Gamma(\ell_i)) \le \delta + \sigma(G_C)(1-\delta) \le \sigma(G_C)$.
\hfill\qed

\bigskip \noindent It was mentioned that the row constraints are also
necessary conditions if and only if the conflict graph is the
disjoint union of complete graphs.  This observation follows as a
special case of
Theorem~\ref{theorem:star:number:row:factor}: the row
constraints which are sufficient conditions are also necessary iff
$P_{\row} = P_I$, which is the case iff $\sigma(G_C)=1$, which is the case iff each
component of $G_C$ is a complete graph.  While the induced star number of a graph can be arbitrarily large, for special classes of networks studied in the literature this quantity is bounded by a fixed constant. This happens to be in the case for unit disk graphs and for networks with primary interference constraints.  We examine these special cases in Section~\ref{section:examples}.

\subsection{A strengthening of the row constraints}

We saw above that the row constraints are a sufficient condition for feasibility of a link demand vector, and the performance of this distributed algorithm is determined by the induced star number $\sigma(G_C)$. We now show that a slight improvement to $\sigma(G_C)-1$ can be obtained.

\bigskip \noindent Recall that the row constraint corresponding to link $\ell_i$ is that the sum total of the demand $\tau(\ell_i)$ and the demands of \emph{all} its interfering neighbors $\tau(\Gamma(\ell_i))$ not exceed the available resource $T$.  It is easy to see that all the links in the network, except for any one designated link, say $\ell_1$, can ignore the demand of up to \emph{one} of its interfering neighbors.

\begin{Proposition}
Given a network and its conflict graph $G_C$, pick any designated link $\ell_1 \in L$.  A sufficient condition for $\tau$ to be feasible within duration $T$ is that
\begin{eqnarray*}
\tau(\ell_i) + \tau(\Gamma(\ell_i))&&\le T,~ i=1 \\
\tau(\ell_i) + \{\tau(\Gamma(\ell_i)) - \min_{\ell_j \in \Gamma(\ell_i)} \tau(\ell_j) \}&&\le T, ~i=2,\ldots,m.
\end{eqnarray*}
\end{Proposition}

\noindent \emph{Proof:}  Suppose the inequalities in the assertion are satisfied.  Order the links, starting with $\ell_1$, as follows.  Let $\ell_2$ be any link adjacent (in $G_C$) to $\ell_1$, let $\ell_3$ be any link adjacent to $\ell_1$ or $\ell_2$.  Given $\ell_1,\ldots,\ell_r$, let $\ell_{r+1}$ be any link adjacent to one of the previous links. An ordering $\ell_1,\ell_2,\ldots,\ell_m$ of all the links is thus obtained. The scheduling mechanism assigns time intervals to these links in reverse order, starting with $\ell_m$.   Assign the time interval $[0,\tau(\ell_m)]$ to $\ell_m$.  Once the links $\ell_m,\ell_{m-1},\ldots,\ell_{r+1}$ have been scheduled, by the inequality above for $\ell_r$ the demand $\tau(\ell_r)$ can also be satisfied because  $\ell_r$  has at least one neighbor in $\{\ell_1,\ldots,\ell_{r-1}\}$ which has not yet been scheduled.  Finally, $\ell_1$ can also be scheduled because of the inequality above for $\ell_1$.
\hfill\qed

\bigskip \noindent Note that this sufficient condition is equivalent to the row constraints when $G_C$ is complete.  This is because the inequality for $i=1$ becomes $\tau(L) \le T$, which implies that the set of inequalities for $i=2,\ldots,m$ are also satisfied. If $G_C$ is not complete, then it is possible to do away with the notion of a designated link (i.e. \emph{every} link can ignore the demand of up to one of its neighbors), provided the conflict graph is not of one exceptional type - the odd cycle.  The proof is by reducing to the previous case by showing that the scheduling mechanism can always find a designated link for which the conditions above are essentially satisfied.

\begin{Proposition}
Suppose $G_C$ is not complete.  Then the set of constraints
$$\tau(\ell_i) + \{\tau(\Gamma(\ell_i)) - \min_{\ell_j \in \Gamma(\ell_i)} \tau(\ell_j) \} ~\le ~T,~~~i=1,\ldots,m $$
is a sufficient condition for $\tau$ to be feasible within duration $T$ if $G_C$ is not an odd cycle.
Furthermore, the smallest scaling factor that converts this sufficient condition into a necessary one is equal to exactly $\sigma(G_C)$ or $\sigma(G_C)-1$, depending on the structure of $G_C$.
\end{Proposition}

\noindent \emph{Proof:} Suppose the inequalities in the assertion are satisfied.
Let $r \ge 1$ be the minimum number of vertices of $G_C$ whose removal disconnects $G_C$ into more than one connected component. We consider three cases, depending on the value of $r$ (this proof method is from \cite{Brooks:1941}):

\emph{$r=1$}:  Let $\ell_1$ be a cut-vertex of $G_C$, so that the removal of $\ell_1$ produces $s \ge 2$ connected components $G_1,\ldots,G_s$.  Since each $G_i$ is connected to $\ell_1$ and by the inequalities of the condition,  the demand of each link in $L - \{\ell_1 \}$ can be satisfied, using the scheduling mechanism given in an earlier proof and using $\ell_1$ as the designated link.  It remains to show that the demand of $\ell_1$ can also be satisfied.  Indeed, note that $\ell_1$ has neighbors $\ell_a$ and $\ell_b$ in $G_1$ and $G_2$, respectively, where $\ell_a$ and $\ell_b$ are nonadjacent.  Hence, the time slots of the schedule in $G_1$ can be permuted so that the time interval assigned to $\ell_a$ is a subset of that assigned to $\ell_b$ (or, in case $\tau(\ell_a) > \tau(\ell_b)$, the time slots can be permuted so that the time interval assigned to $\ell_b$ is a subset of that assigned to $\ell_a$).  Now $\ell_1$ can be scheduled as well since two of its neighbors have been assigned overlapping intervals.

\emph{$r \ge 3$}: Since $G_C$ is not complete, there exist $\ell_1$, $\ell_2$ and  $\ell_3$ such that $\ell_1$ is adjacent to both $\ell_2$ and $\ell_3$ while $\ell_2$ and $\ell_3$ are nonadjacent.  Since $r \ge 3$, $G_C - \{ \ell_2,\ell_3\}$ is connected.  First assign $\ell_2$ and $\ell_3$ the time intervals $[0,\tau(\ell_2)]$ and $[0,\tau(\ell_3)]$, respectively.  By the inequalities of the condition, we can now schedule the remaining links $\ell_4,\ldots,\ell_m$ in some order, with $\ell_1$ as the designated link of the connected graph $G_C - \{\ell_2,\ell_3 \}$.  Finally, $\ell_1$ can also be scheduled since two of its neighbors have been assigned overlapping intervals.

\emph{$r = 2$}: Let $\Delta$ denote the maximum degree of a vertex of $G_C$.  If $\Delta \le 2$, then the graph is either an odd cycle or an even cycle.  If $G_C$ is an even cycle, say $(\ell_1,\ldots,\ell_{2k})$, then a feasible schedule is obtained by assigning to each link of odd index $i$ the time interval $[0,\tau(\ell_i)]$ and to each link of even index $i$ the time interval $[T-\tau(\ell_i),T]$.  If $G_C$ is an odd cycle, say $(\ell_1,\ldots,\ell_{2k+1})$, note that the demand vector $(\tau(\ell_i)=1/2, \forall \ell_i \in L)$ satisfies the inequalities of the condition but is not feasible within duration $T$. So now assume $\Delta \ge 3$.   There exist $\ell_1$ and $\ell_2$ such that $\ell_1$ is a cut-vertex of $G_C - \{\ell_2\}$. So suppose the removal of this cut-vertex decomposes $G_C - \{\ell_1,\ell_2 \}$ into connected components $G_1,\ldots,G_s$.  Now $\ell_1$ has neighbors $\ell_a$ and $\ell_b$ in $G_1$ and $G_2$, respectively.  Note that $\ell_a$ and $\ell_b$ are nonadjacent and $G_C - \{\ell_a,\ell_b \}$ is connected.  As before,  we can assign time intervals $[0,\tau(\ell_a)]$ and $[0,\tau(\ell_b)]$ to $\ell_a$ and $\ell_b$, respectively, then schedule all the remaining links with $\ell_1$ as the designated link, and finally schedule $\ell_1$ as well.

This proves the sufficiency of the condition.

We now determine the performance $\beta_{\row2}$ of this distributed algorithm, where $\beta_{\row2}$ denotes the smallest scaling factor which converts this sufficient condition into a necessary one, i.e., replacing $T$ by $\beta_{\row2} T$ in the inequalities above produces a set of necessary conditions.  Let $\sigma$ denote $\sigma(G_C)$.  Now $G_C$ contains the star graph on some nodes $\ell_1,\ldots,\ell_{\sigma+1}$ as an induced subgraph, with $\ell_1$ as the center node.  For these nodes, the demand $(\varepsilon,1-\varepsilon,\ldots,1-\varepsilon)$ is feasible within one unit of time, and the left-hand-side of the inequality for $\ell_1$ given in the sufficient condition evaluates to $\varepsilon + (\sigma-1)(1-\varepsilon)$, which can be made arbitrarily close to $\sigma-1$.  Hence, $\beta_{\row2} \ge \sigma-1$.  Also, since this sufficient condition is stronger than the row constraints, $\beta_{\row2} \le \beta_{\row} = \sigma$.  So we have that  $\sigma-1 \le \beta_{\row2} \le \sigma$.

Let $S \subseteq L$ denote the vertices $\ell$ of $G_C$ that can induce a maximum sized star of $G_C$ with some of their neighbors, i.e. $S:=\{\ell \in L: \alpha(G_C[\Gamma(\ell)])=\sigma\}$; here $\alpha(G)$ denotes the maximum size of an independent set of $G$.  (It suffices to consider just these vertices because the remaining vertices have a smaller value of $\alpha(G_C[\Gamma(\ell)])$, and hence the left hand side of the inequalities in the condition evaluate to at most $\sigma$.)  If the degree of every vertex in $S$ is exactly $\sigma$, then $\beta_{\row2} = \sigma-1$, as we just showed.  However, suppose the degree of some vertex $\ell'$ in $S$ is at least $\sigma+1$, where $\{\ell_1,\ldots,\ell_{\sigma+1}\}$ are the neighbors of $\ell'$ and the first $\sigma$ of these neighbors form an independent set.  Then the demand vector for $(\ell',\ell_1,\ldots,\ell_{\sigma+1})$ given by $(\varepsilon,1-\varepsilon-\varepsilon/2,\ldots,1-\varepsilon-\varepsilon/2, \varepsilon/2)$ is feasible within one unit of time, but the left-hand-side of the inequality for $\ell'$ given in the condition can be made arbitrarily close to $\sigma$.  Hence $\beta_{\row2} \ge \sigma$ in this case.  It follows that $\beta_{\row2}$ equals  $\sigma$ or $\sigma-1$, according as whether the degree of any vertex in $S$ exceeds or doesn't exceed $\sigma$.
\hfill\qed

\section{Degree and mixed conditions}
\label{sec:degree:mixed}
It was shown that the row constraints provided a simple, distributed sufficient condition for feasibility of a given demand vector. In this condition, there was exactly one constraint associated with each link, namely, the sum total of the demand of the link and demands of its neighbors not exceed the available resource. We now describe an even simpler condition.  We call this the degree condition since it requires knowing, for each link, the demand of that link and just the \emph{number} (not actual demands) of links interfering with it.

\bigskip \noindent Suppose link $\ell_i$ interferes with exactly $d(\ell_i)$ other links, i.e. in the conflict graph $\ell_i$ has degree $d(\ell_i)$.  Then, the following result provides another sufficient condition for admission control \cite{Hamdaoui:Ramanathan:05}:
\begin{Proposition}
\label{proposition:degree:condition:sufficient}
A given demand vector $(\tau(\ell_i): \ell_i \in L)$ is feasible within duration $T$ if $\tau(\ell_i) (d(\ell_i)+1) \le T$ for each $\ell_i \in L$.
\end{Proposition}

\noindent \emph{Proof:}
Suppose the inequalities in the assertion are satisfied. Order the links so that $\tau(\ell_1) \le \tau(\ell_2) \le \ldots \le \tau(\ell_m)$.  Assign $\ell_1$ the time interval $[0,\tau(\ell_1)]$.  Assume links $\ell_1,\ldots,\ell_r$ have already been assigned time intervals satisfying their demands.  It follows from $\tau(\ell_{r+1}) (d(\ell_{r+1})+1) \le T$ and the chosen ordering of the links that
$\tau(\ell_{r+1}) + \tau(\Gamma(\ell_{r+1}) \cap \{\ell_1,\ldots,\ell_r \}) \le T$.  Hence, it is possible to assign to $\ell_{r+1}$ a subset of $[0,T]$ of length $\tau(\ell_{r+1})$ which is disjoint from the intervals already assigned to its neighbors.  It follows by induction that the demand vector is feasible.
\hfill\qed

\bigskip \noindent The degree condition and the row constraints are seen to be equivalent when all links have the same demand value.  Also, note that while the admission control mechanism specified by the degree condition can be implemented in a distributed manner, the scheduling mechanism given in the proof above requires ordering all the links according to their demand values, which is global information.  We will see more examples of such conditions where the admission control mechanism is distributed but the scheduling mechanism realizing a feasible schedule is not distributed.

\bigskip \noindent The degree condition is quite loose (far away from optimal) in that it admits a given demand vector $\tau$ only if $\tau(\ell_i) \le 0.5~T$ for each $\ell_i$ (this is because, since $G_C$ is connected, we have $d(\ell_i) \ge 1$), and more generally, only if $\tau(\ell_i) \le T/(d(\ell_i)+1)$ for each $\ell_i \in L$.  In particular, when the demand values of the various links are asymmetric - in the sense that some links have small demand values, while others have demand values close to the maximum possible $T$ - then the demand vector will not be admitted by the degree condition even if it really is feasible.


\bigskip \noindent The performance of the degree condition is determined, not surprisingly, by the maximum degree of a vertex in the conflict graph.  More precisely, define
$$P_{\degree} := \{\tau \ge 0: \tau(\ell_i)(d(\ell_i)+1) \le 1,~ \forall i\}.$$
Then $P_{\degree} \subseteq P_I$ by Proposition~\ref{proposition:degree:condition:sufficient}.   Define $$\beta_{\degree} := \inf\{\beta \ge 1: P_I \subseteq \beta P_{\degree} \}.$$  Let $\Delta(G_C)$ denote the maximum degree of a vertex in $G_C$.

\begin{Lemma}
For any conflict graph $G_C$,  the exact worst-case performance of the degree condition is given by  $\beta_{\degree}(G_C) = \Delta(G_C)+1$.
\end{Lemma}

\noindent \emph{Proof:}
Let $\ell_1$ be a link having exactly $d$ neighbors $\ell_2,\ldots,\ell_{d+1}$ in the conflict graph.  For these $d+1$ links, the demand vector $\tau=(1-\varepsilon,\varepsilon/d,\ldots,\varepsilon/d)$ is feasible within one unit of time, and $\tau(\ell_1) (d+1) = d+1-\varepsilon(d+1)$ can be made arbitrarily close to $d+1$.  Hence, $\beta_{\degree}(G_C) \ge \Delta(G_C)+1$.

Now suppose $\tau$ is feasible within one unit of time, and let $\ell_i$ be any link.  Then, $\tau(\ell_i) \le 1$, and so $\tau(\ell_i)(d(\ell_i)+1) \le 1(\Delta(G_C)+1)$.  Hence, $\tau \in (\Delta(G_C)+1)P_{\degree}$.
\hfill\qed

\bigskip \noindent This implies that the degree condition is also necessary (and hence optimal) iff $G_C$ is the empty graph, i.e. iff no two links interfere with each other.  It is possible to combine the row constraints and degree constraints to get a sufficient condition which is strictly stronger, as shown in \cite{Hamdaoui:Ramanathan:05}:

\begin{Proposition}
A link demand vector $\tau$ is feasible within duration $T$ if
$$ \min \{~\tau(\ell_i)+\tau(\Gamma(\ell_i))~,~\tau(\ell_i)(d(\ell_i)+1) ~\} \le T, ~\forall \ell_i \in L.$$
\end{Proposition}

\noindent \emph{Proof:} Suppose the inequalities in the condition are satisfied.  Order the links so that $\tau(\ell_1) \le \ldots \le \tau(\ell_m)$.  Assign $\ell_1$ the time interval $[0,\tau(\ell_1)]$.  Suppose $\ell_1,\ldots,\ell_r$ have already been assigned time intervals satisfying their demands. By the inequality in the condition for $\ell_{r+1}$, either $\tau(\ell_i)+\tau(\Gamma(\ell_i)) \le T$ or $\tau(\ell_i)(d(\ell_i)+1) \le T$; in either case, $\ell_{r+1}$ can also be scheduled, as was shown in the earlier proofs where one of these conditions is satisfied.
\hfill\qed

\bigskip \noindent  As introduced in \cite{Hamdaoui:Ramanathan:05}, we call this sufficient condition the \emph{mixed condition}.
Consider the conflict graph $G_C$ on the three links $\{\ell_1,\ell_2,\ell_3\}$, where $\ell_1$ is adjacent to each of $\ell_2$ and $\ell_3$, and $\ell_2$ and $\ell_3$ are nonadjacent. Let $T=1$.  Note that the link demand vector $(0.9,0.1,0.1)$ satisfies the row constraints but not the degree constraints, and the demand $(0.3,0.5,0.5)$ satisfies the degree constraints but not the row constraints.  Hence these two conditions are incomparable.  Also, the demand $(0.3,0.6.0.6)$ satisfies neither the row nor the degree constraints but does satisfy the mixed condition. Finally, the demand $(1/3 + \varepsilon, 1/3, 1/3)$ is feasible but does not satisfy the mixed condition.  (This last example proves that the mixed condition is optimal (and hence, also necessary) iff $G_C$ is the disjoint union of complete graphs, i.e. iff $\sigma(G_C)=1$, an observation which we can also deduce from a more general result; cf. Theorem~\ref{theorem:mixed:condition:performance}.)  These examples show that, in general,
$$P_{\row},P_{\degree}~ \varsubsetneq ~(P_{\row} ~\cup~ P_{\degree}) ~\varsubsetneq ~P_{\mixed} ~\varsubsetneq~ P_I,$$
where
$$P_{\mixed} := \{\tau: \min \{~\tau(\ell_i)+\tau(\Gamma(\ell_i))~,~\tau(\ell_i)(d(\ell_i)+1) ~\} \le 1,~\forall \ell_i \in L \}.$$

\bigskip \noindent
Let $\beta_{\mixed}$ denote the smallest scaling factor that converts the sufficient mixed condition into a necessary one; hence, given the conflict graph $G_C=(L,L')$ and its independent set polytope $P_I$, we have that $P_{\mixed} \subseteq P_I \subseteq \beta_{\mixed} P_{\mixed}$ and
$$ \beta_{\mixed} = \sup_{\tau \in P_I}~ \max_{\ell_i \in L}~ \min\{~\tau(\ell_i)+\tau(\Gamma(\ell_i))~,~\tau(\ell_i)(d(\ell_i)+1) ~ \} .$$

\begin{Theorem}
\label{theorem:mixed:condition:performance}
The worst-case performance of the mixed condition is bounded as
$$ \frac{1+\sigma(G_C)}{2} ~\le ~ \beta_{\mixed} ~ \le ~ \sigma(G_C),$$
where $\sigma(G_C)$ denotes the induced star number of $G_C$.  Moreover, the lower and upper bounds are tight; the star graphs realize the lower bound, and there exist graph sequences for which $\beta_{\mixed}$ approaches the upper bound arbitrarily closely.
\end{Theorem}

\noindent \emph{Proof:}  Since the mixed condition is satisfied whenever the row constraints are satisfied, $\beta_{\mixed} \le \beta_{\row}$, and so the upper bound follows from Theorem~\ref{theorem:star:number:row:factor}.

To prove the lower bound, suppose $G_C=(L,L')$ is a star graph on vertices $L=(x,y_1,\ldots,y_{\eta})$, where $x$ is the center vertex and the leaf vertices $y_i$ form an independent set.  We show that for this $G_C$ and its independent set polytope $P_I$,
$$\sup_{\tau \in P_I}~\max_{\ell_i \in L}~ \min\{~\tau(\ell_i)+\tau(\Gamma(\ell_i))~,~\tau(\ell_i)(d(\ell_i)+1) ~ \} = \frac{1+\eta}{2}.$$
It can be assumed that $\eta \ge 2$, for if $\eta=1$ then $G_C$ is complete and so $\beta_{\mixed}=1$ by the optimality of the mixed condition in this case.  Set $\tau(x) =: \delta \in [0,1]$.  In order to maximize
$$\max_{\ell_i \in L}~ \min\{~\tau(\ell_i)+\tau(\Gamma(\ell_i))~,~\tau(\ell_i)(d(\ell_i)+1) ~ \}$$ over $\tau \in P_I$,  $\tau(y_i)$ can be set to equal its maximum possible value of $1 - \delta$, for $i=1,\ldots,\eta$. The mixed condition evaluated at $y_i \in L$ is bounded as $$\min\{~\tau(\ell_i)+\tau(\Gamma(\ell_i))~,~\tau(\ell_i)(d(\ell_i)+1) ~ \} \le \tau(\ell_i)+\tau(\Gamma(\ell_i)) = (1-\delta)+\delta = 1.$$  But $\beta_{\mixed} \ge 1$ trivially.  Hence, to determine the worst-case performance of the mixed condition, it suffices to evaluate the mixed condition at just $x \in L$:
\begin{eqnarray*}
\beta_{\mixed} &=& \sup_{\delta \in [0,1]}~\min\{~\tau(x)+\tau(\Gamma(x))~,~\tau(x)(d(x)+1) ~\} \\
&=& \sup_{\delta \in [0,1]}~\min\{\delta+\eta(1-\delta)~,~\delta(1+\eta)\}.
\end{eqnarray*}
Note that $\delta+\eta(1-\delta) \le \delta(1+\eta)$ iff $\delta \ge 1/2$.  It can be seen that the optimum is attained at $\delta = 1/2$, giving $\beta_{\mixed} = \frac{1+\eta}{2}$ for this star graph on $1+\eta$ vertices.

This proves the lower bound in the assertion, and it has also been shown that the class of star graphs realize this lower bound.

We now construct a sequence of graphs for which $\beta_{\mixed}$ approaches the upper bound in the limit.  We do this by determining the exact value of $\beta_{\mixed}$ for a wider class of graphs which includes the class of star graphs as a special case.  The property that $G_C=(L,L')$ needs to satisfy is that it has some $x \in L$ that is adjacent to every other element of $L$ and such that the removal of $x$ disconnects $G_C$ into a disjoint union of complete graphs, i.e., $x$ has degree $|L|-1$ and $\sigma(G_C - x) \le1$.

We now claim the following: Suppose $G_C=(L,L')$ is such that $x \in L$ is adjacent to all other members of $L$ and the removal of $x$ produces disjoint complete graphs on vertex sets $L_1,\ldots,L_{\eta}$.  Then
$$\beta_{\mixed} = \frac{\eta(1+\sum|L_i|)}{\eta+\sum|L_i|}.$$  To prove this claim, suppose $G_C=(L,L')$, $x \in L$, and $L_1,\ldots,L_{\eta}$ satisfy the conditions of the claim.  Recall that
$$ \beta_{\mixed} = \sup_{\tau \in P_I}~ \max_{\ell_i \in L}~ \min\{~\tau(\ell_i)+\tau(\Gamma(\ell_i))~,~\tau(\ell_i)(d(\ell_i)+1) ~ \} .$$
Let $\tau(x) =: \delta \in [0,1]$. In order to maximize
$$\max_{\ell_i \in L}~ \min\{~\tau(\ell_i)+\tau(\Gamma(\ell_i))~,~\tau(\ell_i)(d(\ell_i)+1) ~ \}$$
over $\tau \in P_I$, assign demands to the elements of $L_1,\ldots,L_{\eta}$ arbitrarily so that $\tau(L_i) = 1-\delta, i=1,\ldots,\eta$. For this $\tau$, the row constraint at each $y \in L_i$ evaluates to exactly 1 and the mixed condition at each $y$ evaluates to at most 1. Since $\beta_{\mixed} \ge 1$,  $\beta_{\mixed}$ is determined by the value of the mixed condition at just $x$:
\begin{eqnarray*}
\beta_{\mixed} &=& \sup_{\delta \in [0,1]}~\min\{~\tau(x)+\tau(\Gamma(x))~,~\tau(x)(d(x)+1) ~\} \\
& = &\sup_{\delta \in [0,1]}~\min\{~\delta+(1-\delta)\eta~,~\delta(1+\sum|L_i|)~\}.
\end{eqnarray*}
It can be verified that $\delta+(1-\delta)\eta \le \delta(1+\sum|L_i|) $ iff $\delta \ge \frac{\eta}{\eta+\sum|L_i|}$ and that $\beta_{\mixed}$ attains its optimal value of $\frac{\eta(1+\sum|L_i|)}{\eta+\sum|L_i|}$ when $\delta = \frac{\eta}{\eta+\sum|L_i|}$.  This proves the claim.

Note that the case of $|L_i|=1,~\forall~i$, makes $G_C$ a star graph, whereas letting $|L_1|$ approach infinity gives that $\beta_{\mixed}$ approaches the upper bound of $\sigma(G_C)$.   This proves that the upper bound in the assertion is tight.
\hfill\qed

\bigskip \noindent One general class of graphs that includes the star graphs, the even and odd cycles, the complete graphs and bipartite graphs are those that satisfy the following property: for each vertex $\ell \in L$ in the graph $G_C=(L,L')$, the neighbors of $\ell$ induce a disjoint union of complete graphs.  For this general class of graphs there is a simple expression for the exact value of $\beta_{\mixed}$:

\begin{Theorem}
\label{theorem:mixed:performance:exact}
Suppose $G_C=(L,L')$ satisfies $\sigma(G_C[\Gamma(\ell)]) \le 1, ~\forall \ell \in L$.   Let $d(\ell)$ denote the number of neighbors of $\ell$ and let $\eta_{\ell}$ denote the number of connected components induced by the neighbors of $\ell$.  Then
$$\beta_{\mixed} = \max_{\ell \in L} \frac{\eta_{\ell} (1+d(\ell))}{\eta_{\ell} + d(\ell)}.$$
\end{Theorem}

This result implies that if $G_C$ is a star graph, a bipartite graph, or a cycle graph, then the exact worst-case performance of the mixed condition is $\beta_{\mixed} = \frac{1+\Delta(G_C)}{2}$.  This value is about a half of $\Delta(G_C)$ or $\sigma(G_C)$ for some families of graphs; hence,  in terms of the amount of resources requested, it is seen that the mixed condition can be an improvement over the row constraints and the degree condition by up to a factor of 1/2.

The proof of Theorem~\ref{theorem:mixed:performance:exact} is similar to the proof of the claim in the previous proof and so the details are omitted.  The simplest example of a graph that does not satisfy the conditions of Theorem~\ref{theorem:mixed:performance:exact} is the graph $K_4 - e$ (one edge removed from the complete graph on 4 vertices).  For this graph, Theorem~\ref{theorem:mixed:condition:performance} immediately gives that the graph invariant is bounded as $1.5 \le \beta_{\mixed} \le 2$. A straightforward but lengthy computation yields the exact value of $\beta_{\mixed}=1.6.$

\section{Clique constraints}
\label{sec:clique:constraints}
A necessary condition for a given link demand vector to be feasible can be obtained as follows. Suppose there exists a schedule of duration 1 satisfying demand $\tau$.  Then if $K$ is a clique in the conflict graph, the time intervals assigned to the distinct links in $K$ must be disjoint, hence $\tau(K) \le 1$.  Thus, a necessary condition for $\tau$ to be feasible within duration $T$ is that $\tau(K) \le T$ for every maximal clique $K$ in the conflict graph. These constraints are called \emph{clique constraints} \cite{Gupta:Musacchio:Walrand:07}.  As before, we can associate a polytope with this necessary condition; define
$$P_{\clique} := \{\tau: \tau(K) \le 1, \forall  K \}  \supseteq P_I,$$  where $K$ runs over all the cliques (or equivalently, over just all the maximal cliques) of the conflict graph.

\bigskip \noindent A perfect graph is one for which every induced subgraph has its chromatic number equal to its clique number. It is known that $P_I = P_{\clique}$ for a graph if and only if the graph is perfect.
%
%
%
%

\bigskip \noindent
Using the notion of the imperfection ratio of graphs, bounds on the suboptimality of clique constraints were obtained \cite{Gupta:Musacchio:Walrand:07} for the case of unit disk graphs.  More precisely, given a conflict graph $G_C$ and demand vector $\tau$, let $T^*(\tau)$ denote the minimum duration of a schedule satisfying $\tau$ (the optimal value of this linear program is also the smallest $\beta$ such that $\tau \in \beta P_I$, and $T^*(\textbf{1})$ is often referred to as the fractional chromatic number of $G_C$). Let $T_{\clique}(\tau)$ denote the maximum value of $\tau(K)$ over all cliques $K$ in the conflict graph; so $T_{\clique}(\tau) \le T^*(\tau)$.  The imperfection ratio of a graph $G_C$ is defined as
$$\imp(G_C) := \sup_{\tau \ne 0} \{ T^*(\tau) / T_{\clique}(\tau) \} .$$
This quantity has been studied in \cite{Gerke:McDiarmid:2001}; it is finite and is achieved for any given graph.  In the definition above, for a given demand vector $\tau$, the numerator specifies the exact amount of resource required to satisfy the demand, as determined by an optimal, centralized algorithm.  The denominator specifies a lower bound on the resource required to satisfy the demand, as determined by a particular distributed algorithm (the clique constraints).  Their ratio is the factor by which the distributed algorithm is away from optimal for the given demand vector.  The imperfection ratio, which maximizes this ratio over all demand patterns, is then the worst-case performance of the distributed algorithm.  This argument is made more precise in the next proof.

The following general result is implicit in \cite{Gupta:Musacchio:Walrand:07} (where the authors focus on unit disk graphs) and in \cite{Gerke:McDiarmid:2001}:

\begin{Proposition}
\label{prop:clique:constraints:imp}
The largest scaling factor which converts the necessary clique constraints into a sufficient condition is $1 / \imp(G_C)$; i.e. the worst-case performance of the clique constraints is given by
$$\sup\{\beta \le 1: \beta P_{\clique} \subseteq P_I \} = \frac{1}{\imp(G_C)};$$
and
$$ \frac{1}{\imp(G_C)} P_{\clique} \subseteq P_I \subseteq P_{\clique} .$$
\end{Proposition}

\noindent \emph{Proof:}
Suppose $\tau \in \frac{1}{\imp(G_C)} P_{\clique}$. Then $\tau(K) \le \frac{1}{\imp(G_C)}$ for all cliques $K$.  So $T^*(\tau) \le \imp(G_C) T_{\clique}(\tau) \le \imp(G_C) \frac{1}{\imp(G_C)} \le 1$. So $\tau \in P_I$.

Now suppose $\beta > 1/\imp(G_C)$. It suffices to show that $\beta P_{\clique} \nsubseteq P_I$.  Since $\imp(G_C) > 1/\beta$, there exists a $\tau$ such that $T^*(\tau) / T_{\clique}(\tau) > 1/\beta$.  Define $\tilde{\tau} := \frac{\beta}{T_{\clique}(\tau)} \tau $. Then $\tilde{\tau} \in \beta P_{\clique}$, but $T^*(\tilde{\tau}) = \frac{\beta}{T_{\clique}(\tau)} T^*(\tau) > 1$.
\hfill\qed

The problem of determining the imperfection ratio of various families of graphs has been studied in \cite{Gerke:McDiarmid:2001}.
\section{Examples}
\label{section:examples}

In this section we apply the results obtained so far to some special classes of networks and interference models.   In Section~\ref{section:primary:interference}, we examine a model of interference called primary interference constraints, which has been well-studied in the literature in the centralized setting; we examine we examine this problem in the distributed setting. In Section~\ref{section:unit:disk:networks} we look at the unit disk model, which is quite popular and widely used by researchers in the sensor networks community to model the topology of a network.

\subsection{Primary interference model}
\label{section:primary:interference}
Given a network $G=(V,L)$, suppose two links are considered to be interfering iff they share one or more endvertices in common.  We refer to this kind of interference as primary interference.   This interference model arises, for example, from the assumption that each node can communicate to only one other node at any given time.
This interference model is perhaps the most well studied in the literature; for example, see \cite{Hajek:1984}, \cite{Hajek:Sasaki:1988},  \cite{Kodialam:Nandagopal:2005}.  The conflict graph for such a network is called a \emph{line graph}.
 Note that in this kind of interference model, if link $\ell_1$ interferes with each of $\ell_2,\ell_3$ and $\ell_4$, and $\ell_2$ doesn't interfere with $\ell_3$, then $\ell_4$ must interfere with exactly one of $\ell_2$ or $\ell_3$.  So we obtain a well-known result that if the conflict graph $G_C$ is a line graph then $\sigma(G_C) \le 2$.  It follows that for such networks the row constraints will be at most a factor 2 away from optimal.

\bigskip \noindent More specifically, for this interference model, the row constraints on the conflict graph can be reformulated on the network graph $G=(V,L)$ as follows.  Suppose link $\ell_i$ is incident between nodes $u_i$ and $v_i$.  Given link demand vector $\tau$, let $\tau(u)$ denote the sum of the demands of all links incident in $G$ to node $u$. Then the row constraint $\tau(\ell_i) + \tau(\Gamma(\ell_i)) \le T$ in the conflict graph $G_C=(L,L')$ is equivalent to the constraint $\tau(u_i) + \tau(v_i) - \tau(\ell_i) \le T$ in the network graph. This equivalence yields the following sufficient condition:
\begin{Corollary}
\label{corollary:row:constraints:line:graph}
Let $G=(V,L)$ be a network graph, and suppose two links interfere with each other if and only if they are incident to a common node.  Then $(\tau(\ell):\ell \in L)$ is feasible within duration $T$ if for each $\ell=\{u,v\}$, $\tau(u)+\tau(v)-\tau(\ell) \le T$.  This sufficient condition is a factor of at most 2 away from optimal. \hfill\qed
\end{Corollary}

\bigskip \noindent \emph{Remark.} In the case of primary interference constraints, upper bounds on the minimum scheduling time can be obtained immediately by using upper bounds on the chromatic index of multigraphs;  for example, see some of the results in \cite{Hajek:1984},  \cite{Kodialam:Nandagopal:2005}.  We note here a reverse implication: the row constraints above (and its proof) can be used to obtain an upperbound on the chromatic index of multigraphs.  More precisely, let $G=(V,L)$ be a simple undirected graph. Let $\mu\{u,v\}$ denote the number of parallel edges between vertices $u$ and $v$. This defines a multigraph $(G,\mu)$. A proper edge-coloring is an assignment of colors to edges so that any two edges that share one or more endvertices in common are assigned different colors. The chromatic index of a multigraph is the minimum number of colors required to properly color its edges.  Let $\mu(u)$ denote the number of edges incident to $u \in V$.
Then, Corollary~\ref{corollary:row:constraints:line:graph} implies that: the chromatic index of the multigraph $(G,\mu)$ is bounded from above by
$ \max_{\ell=\{u,v\}} (\mu(u) + \mu(v) - \mu\{u,v\}).$


\bigskip \noindent Another distributed algorithm that can be used in networks having primary interference is given by the clique constraints.  Trivially, a necessary condition for $\tau$ to be feasible within duration $T$ is that $\tau(K) \le T$ for all maximal cliques $K$ in the conflict graph.  By Proposition~\ref{prop:clique:constraints:imp}, the performance of the clique constraints is determined by the imperfection ratio of the conflict graph.  It is known that the imperfection ratio of a line graph is at most 1.25 \cite[Prop.~3.8]{Gerke:McDiarmid:2001}.  This means that a sufficient condition for $\tau$ to be feasible within duration $T$ is that $1.25 \tau(K) \le T$ for all maximal cliques $K$ in the conflict graph.  Since $G_C$ is a line graph, each maximal clique $K$ in $G_C$ corresponds either to a set of links $K \subseteq L$ that are all incident to a common node in the network graph $G=(V,L)$ or to a set of three links that form a triangle in the network graph.  For $v \in V$, let $\tau(v)$ denote the sum of the demands of all links incident to $v$ in the network graph.  We have shown that the following result provides an efficient, distributed algorithm for admission control:

\begin{Theorem}
\label{theorem:clique:constraints:line:graph}
Let $G=(V,L)$ be a network graph, and suppose two links interfere with each other if and only if they are incident to a common node. Then $(\tau(\ell): \ell \in L)$ is feasible within duration $T$ if $\tau(v) \le 0.8T,~\forall v \in V$ and $\tau(uv)+\tau(vw)+\tau(uw) \le 0.8T,~\forall u,v,w \in V$. This sufficient condition is a factor of at most 1.25 away from optimal.
\end{Theorem}

\bigskip \noindent An important aspect of this result is that, though the number of maximal cliques in a general graph can grow exponentially with the size of the graph, the number of maximal cliques in a line graph grows only polynomially in the size of the graph.  This is because the maximal cliques in a line graph correspond to one of two different types of cliques, as mentioned above, and there are only a polynomial number of such cliques, as can be verified by just counting the number of possible cliques of these two types by examining the network graph.  Thus, unlike for general networks, for networks with primary interference the clique constraints provide an efficient distributed algorithm for checking feasibility of a given demand vector.

\bigskip \noindent \emph{Remark}:  A result due to Shannon on the edge-coloring of multigraphs \cite{Shannon:1949} implies that: for a given network graph $G=(V,L)$, a sufficient condition for $(\tau(\ell): \ell \in L)$ to be feasible within duration $T$ is that $\tau(v) \le 2T/3,~\forall v \in V$.  Theorem~\ref{theorem:clique:constraints:line:graph} improves this bound from a factor of 2/3 to 0.8.  This improvement is possible because, unlike in the classical edge-coloring problem, fractional coloring solutions are also admissible in our framework.  Furthermore, the sufficient condition in Theorem~\ref{theorem:clique:constraints:line:graph} is less localized, in that each node in the network graph needs to know not only the sum total of the demands of all links incident to it, but also the demands of all links between its neighbors.  



\bigskip \noindent It is possible to construct examples to show that the sufficient conditions provided by Corollary~\ref{corollary:row:constraints:line:graph} and Theorem~\ref{theorem:clique:constraints:line:graph} are incomparable, i.e. the sets $P_{\row}$ and $0.8~P_{\clique}$ are (inclusionwise) incomparable. We omit these details.

\subsection{Unit disk networks}
\label{section:unit:disk:networks}
One assumption that is often used by researchers in the sensor networks community to model the topology of a wireless sensor network is the unit disk assumption.  In this model, all nodes are assumed to have the same transmission power and are equipped with omnidirectional antennas; equivalently, the transmission range for each node is a disk of constant radius centered at that node.  By scaling the axes, the radius can be assumed to be 1, hence the name \emph{unit} disk network for such models.  Thus, in this model two nodes can communicate with each other iff unit disks centered at their locations intersect.  Not all network topologies can be realized with this model: the topology of unit disk networks is very restricted, since if nodes $a$ and $b$ are close to each other, and nodes $b$ and $c$ are close to each other, then $a$ and $c$ are not too far away from each other and hence are also likely to be connected.

\bigskip \noindent A graph is called a unit disk graph if it can be realized as the topology of some unit disk network, i.e. if there exist locations on the Euclidean plane - one location corresponding to each vertex of the graph - such that unit disks centered at a pair of locations intersect if and only if the corresponding vertices of the graph are adjacent.

\bigskip \noindent It is known that the induced
star number of a unit disk graph is at most 5.  To prove this result, suppose some vertex $u$ of a unit disk graph has 6 or more neighbors.  Divide the unit disk region centered at $u$ into six sectors of 60 degrees each.  Then at least two of the vertices adjacent to $u$ must lie in the same 60 degree sector and hence at a distance of at most one unit from each other.  Hence $u$ cannot have 6 or more neighbors that are pairwise nonadjacent.

\bigskip \noindent An example of an interference model for which the conflict graph is a unit disk graph is given in \cite{Gupta:Musacchio:Walrand:07}.  In such cases the following result applies.

\begin{Corollary}
\label{corollary:row:constraints:udg}
If the conflict graph is a unit disk graph, then the row constraints are at most
a bounded factor of 5 away from optimal.
\end{Corollary}

\bigskip \noindent A bound on the performance of another distributed algorithm (the scaled clique constraints) is known for the case where the conflict graph is a unit disk graph.  By Proposition~\ref{prop:clique:constraints:imp}, the performance of the clique constraints is determined by the imperfection ratio of the conflict graph. It is known that the imperfection ratio of a unit disk graph is at most 2.1 \cite[Prop.~3.3]{Gerke:McDiarmid:2001}.  In other words, the clique constraints are necessary conditions (ie. $ P_I \subseteq P_{\clique} $) and scaling the clique constraints by a factor of 1/2.1 gives a sufficient condition for admission control (ie. $0.46 P_{\clique} \subseteq P_I$). This yields the following result \cite[Thm.~2]{Gupta:Musacchio:Walrand:07}:

\begin{Corollary}
\label{corollary:clique:constraints:udg}
Suppose the conflict graph is a unit disk graph.  Then, a link demand vector $\tau$ is feasible within duration $T$ if $\tau(K) \le 0.46~T$ for all cliques $K$ in the conflict graph.
\end{Corollary}

\bigskip \noindent It is possible to construct examples to show that the sufficient conditions provided by Corollary~\ref{corollary:row:constraints:udg} and Corollary~\ref{corollary:clique:constraints:udg} are incomparable, i.e. the sets $P_{\row}$ and $0.46~P_{\clique}$ are (inclusionwise) incomparable. We omit these details.

\section{Fractional chromatic number}
In this section we briefly provide a combinatorial abstraction of some of the results on wireless networks given above.  As we shall see, the results obtained above on the performance of distributed mechanisms for flow admission control in wireless ad-hoc networks can be abstracted and expressed in terms of the performance of certain upper bounds on the fractional chromatic number of weighted graphs. Proposition~\ref{prop:row:constraints:sufficient} and Theorem~\ref{theorem:star:number:row:factor} have been abstracted in this manner and published in \cite{Ganesan:2010}.

Let $G=(V,E)$ be a simple, undirected graph on vertex set $V=\{v_1,v_2,\ldots,v_n\}$.  Let $G_\bx$ be a weighted graph, where $\bx=(x(v_1),\ldots,x(v_n))$ is a nonnegative, real-valued weight assigned to the vertices of $G$.  Let $\{I_1,\ldots,I_L\}$ denote the set of all independent sets of $G$.  Let $B = [b_{ij}]$ be the $n \times L$ vertex-independent set incidence matrix of $G$. Thus, $b_{ij}=1$ if $v_i \in I_j$, and $b_{ij}=0$ otherwise.  Then, the fractional chromatic number of the weighted graph $G_\bx$, denoted by $\chi_f(G_\bx)$,  is defined to the value of the linear program: $\min~\textbf{1}^T \textbf{t}$ subject to $B \textbf{t} \ge \bx$, $\textbf{t} \ge 0$. Equivalently, $\chi_f(G_\bx)$ is the smallest value of $T$ such that each vertex $v_i$ is assigned a subset of $[0,T]$ of total length (or measure) $x(v_i)$ and such that adjacent vertices are assigned subsets that are non-overlapping (except possibly at the endpoints of the intervals).  Thus, given a conflict graph $G$ with a demand vector $\bx$ on its vertices, the problem of estimating upper bounds on the minimum duration of a schedule satisfying these demands using only localized information is essentially the same as that of obtaining upper bounds for $\chi_f(G_\bx)$ using only localized information.

The following two results can now be shown, essentially using the same proof as in Proposition~\ref{prop:row:constraints:sufficient} and Theorem~\ref{theorem:star:number:row:factor}:

\begin{Proposition}
The fractional chromatic number of a weighted graph $G_\bx$ has the upper bound $B(G_\bx)$:
$$ \chi_f(G_\bx) \le \max_{v_i \in V} \left\{x(v_i)+x(\Gamma(v_i)) \right\} =: B(G_\bx).$$
\end{Proposition}
The worst-case performance of this upper bound is the factor by which this upper bound $B(G_\bx)$ is away from the true value $\chi_f(G_\bx)$ for a given weight vector $\bx$, maximized over all possible weight vectors.  This is a new graph invariant
$$\beta(G) :=\sup_{\bx \ne \mathbf{0}} \frac{B(G_\bx)}{\chi_f(G_\bx)},$$
which can be shown to be equal to the induced star number of $G$.  Hence, we have that:
\begin{Theorem} \cite{Ganesan:2010}
For any graph $G$, $$\sup_{\bx \ne \mathbf{0}} \frac{\max_{v_i \in V} \left\{x(v_i)+x(\Gamma(v_i)) \right\}}{\chi_f(G_\bx)} = \sigma(G).$$
\end{Theorem}
The results obtained above on the performance of the other sufficient conditions can also be abstracted in a similar manner to assess the performance of the corresponding stronger upper bounds on the fractional chromatic number; we omit these details.
\section{Concluding remarks}
The exact worst-case performance of some distributed mechanisms for the flow admission problem in wireless ad-hoc networks was characterized.  These performance bounds are important to study because they quantify the extent to which a particular distributed flow admission mechanism can {\em overestimate} the amount of resources required to satisfy a given set of demands.  The results here imply that for some special classes of networks and interference models, there exist simple, efficient, distributed admission control mechanisms and scheduling mechanisms whose performance is within a bounded factor away from that of an optimal, centralized mechanism.   The worst case performance bounds we obtained were simple expressions in terms of graph invariants.

It was seen that the distributed algorithms  discussed here can over-estimate the amount of resources required to carry out a given task by up to a factor equal to the induced star number.  Thus, the performance of distributed systems that employ such resources estimation algorithms is limited by the induced star number of the network.    Hence, when designing such networks, it is desired that this quantity be as close to unity as possible.  It was also seen that for some families of graphs, the mixed condition can be an improvement over the row constraints by up to a factor of 1/2.

It would be worthwhile to extend the results here to more general interference models which cannot be captured by (conflict) graphs but which can be captured by hypergraphs.  Consider, for example, the case of a network consisting of three links that form a triangle, such that up to any two of the three links can be simultaneously active, but if all three links are simultaneously active the interference is not tolerable.  This form of interference can be modeled using a hypergraph, whose hyperedges are the minimial forbidden sets of links (in this particular example, the three links of the triangle would form a hyperedge).  While the hypergraph interference model in wireless communications has been studied in some contexts (cf. \cite{Sarkar:Sivarajan:1998}), topics related to distributed algorithms, fractional chromatic number, upper bounds, and their performance have not yet been investigated for this more general interference model.  Furthermore, if one were to measure and analyze data related to signal to interference and noise ratio, it is possible that the hypergraphs or other combinatorial interference models that arise might have additional structure.

In the primary interference model studied above, it was seen that there was an improvement in the performance of the distributed algorithm from a ratio of 2/3 to 0.8 when the algorithm becomes less localized, i.e. when a node in the network knows not just the sum total of demands of all links incident to it but also the demands of links {\em between its neighbors}.  An important problem is to quantitatively characterize this tradeoff between the level of localization and the performance of the distributed algorithm; one expects that as each node in the network has more global information, the performance of the corresponding distributed algorithm would improve, although there is also a communication cost associated with making information from distant nodes available to a node.


\section*{Acknowledgements}
Thanks are due to professor Parmesh Ramanathan for
 suggesting this direction of scaling the sufficient conditions.
 {
\bibliographystyle{plain}
\bibliography{winet4}
}
\end{document}